\documentclass[10pt]{iopart}
\usepackage[english]{babel}
\usepackage[utf8]{inputenc}
\usepackage{bm}
\usepackage{amsfonts}
\usepackage{iopams}
\usepackage{graphicx}
\usepackage{caption}
\usepackage{subfig}

\begin{document}
\title[Superfluid hydrodynamics of polytropic gases:dimensional reduction and sound velocity]{Superfluid hydrodynamics of polytropic gases: dimensional reduction and sound velocity}
\author{N. Bellomo$^{1}$, G. Mazzarella$^{1,2}$, and L. Salasnich$^{1,2}$} \address{$^{1}$ Dipartimento di Fisica e Astronomia
``Galileo Galilei'', Universit\`a di Padova, Via Marzolo 8, 35131-Padova, Italy\\
$^{2}$
CNISM-Consorzio Interuniversitario per le Scienze Fisiche della Materia}

\begin{abstract}
Motivated by the fact that two-component confined fermionic gases in Bardeen-Cooper-Schrieffer-Bose-Einstein condensate (BCS-BEC) crossover can be described through an hydrodynamical approach, we study these systems - both in the cigar-shaped configuration and in the disk-shaped one - by using a polytropic Lagrangian density. We start from the Popov Lagrangian density and obtain, after a dimensional reduction process, the equations that control the dynamics of such systems. By solving these equations we study the sound velocity as a function of the density by analyzing how the dimensionality affects this velocity.
\end{abstract}

\section{Introduction}
The crossover from the Bardeen-Cooper-Schrieffer (BCS) state of weakly bound Fermi pairs to the Bose-Einstein condensate (BEC) of molecular dimers and systems with reduced dimensionality are two of the most interesting topics in the ultracold atomic physics.

Over the past several years, the predicted BCS-BEC crossover \cite{eagles,leggett,nozieres} has been observed by several experimental groups with $^{6}$Li and $^{40}$K atoms \cite{greiner,regal,kinast,zwierlein,chin,ueda}. To study these systems it is necessary to confine them by suitable traps that, in the simplest case, act as harmonic potentials. Even if harmonically trapped fermions are very dilute systems, their properties are governed by the interaction between particles. This interaction, in the dilute regime and at ultra low temperatures, can be characterized by a single parameter, the s-wave scattering length $a_s$. Experimentally, it is possible to control $a_s$ by using a magnetically tuned Feshbach resonance. This can be described by a simple expression, see \cite{feshbach}, that is $a_s=a_{bg}(1-[\Delta /(B-B_0)])$ with $a_{bg}$ the so called background (off-resonant) s-wave scattering length, $B_0$ the magnetic field strength where the Feshbach resonance occurs, and 
$\Delta$ the resonance width. Then, the Feshbach resonance technique permits one to vary the magnitude and the sign of $a_s$. When $a_s$ is large and negative, the ground state is a BCS superfluid of weakly bound Fermi pairs. For large and positive values of $a_s$, instead, the Bose-Einstein condensation of molecules, associated to the formation of two-body bound states, is observed. In between these two states, there is a smooth crossover when $a_s$ changes its sign by passing through $\pm \infty$ corresponding to $B$ at the resonance position (see the above reported formula). At zero temperature and only in the case of dilute fermionic systems the only relevant dimensionless parameter in the description of the physical properties is $y=(k_Fa_s)^{-1}$, where $k_F$ is the Fermi wave vector.

From the experimental point of view, it is possible to change the harmonic frequencies  inducing a dimensional reduction of the system \cite{gorlitz,schreck,kinoshita}: from three dimensions (3D) to one dimension (1D) and from 3D to two dimensions (2D). The first case takes place with an external trapping realized by superimposing a strong harmonic confinement in the transverse radial plane to a generic shallow potential in the axial direction (cigar-shaped configuration). The second circumstance occurs, instead, when the atoms are trapped by a strong harmonic potential in the axial direction plus a loose potential in the transverse plane (disk-shaped configuration). Both the effects of reduced dimensions \cite{olshanii,jackson,chiofalo,petrov} and the transition from a dimensional regime to another have been carefully studied \cite{salasnich1,salasnich2}. In particular, it has been suggested that a reduced dimensionality strongly modifies density profiles \cite{schneider,sala0,sala1,vignolo}, collective modes \cite{bruun,minguzzi} and stability of mixtures \cite{das,sala2}. Also sound velocities have been theoretically investigated in reduced dimensions for both normal \cite{minguzzi,capuzzi1,suonozero} and superfluid Fermi gases \cite{ghosh,capuzzi2,capuzzi3}. Very recently, Shanenko and co-workers have carried out an interesting analysis about BCS-BEC crossover induced by quantum-size effects in the cigar-shaped configuration \cite{shanenko}.

Both in the cigar-shaped case and in the disk-shaped one the dynamics is practically frozen in the directions where the harmonic trap has been generated. The idea is therefore to describe the system by eliminating the frozen spatial variables by performing a dimensional reduction process as done in \cite{salasnich1,salasnich2}. Salasnich and co-workers have studied the problem by achieving an effective one (two) dimensional wave equation, that is a time-dependent 1D(2D) nonpolynomial Schr\"odinger equation, by integrating out the directions where the dynamics is blocked \cite{salasnich1}. In particular, in Ref. \cite{salasnich2} the authors have studied the sound velocity for Bose-condensed alkali-metal vapors. 

For fermionic systems it would be, in principle, to follow a similar approach, i.e. to use a suitable generalization to the fermionic case of the Gross-Pitaevskii equation (see, for example, Refs. \cite{salasnichnse1,salasnichnse2}). However, here, we study a confined zero-temperature two-component Fermi gas both in cigar-shaped configuration and in the disk-shaped one by approaching the problem from another perspective. We perform the reductional dimensional process by starting from the Popov Lagrangian (PL) density \cite{popov}. This Lagrangian density- via the internal energy of the system - embeddies both the possibility to cross the boundary between the BCS side and the BEC region of the crossover - by means of a polytropic state equation \cite{politropic} - and the effects of the quantum pressure by a gradient correction \cite{weiz,kirzhnits,holas,salasnich3,salasnichnse1}. For the density $n$, appearing in the PL, we use an ansatz which consists in factorizing $n$ as a Gaussian function \cite{salasnich1} in the radial (axial) directions times a generic function in the others. By employing this ansatz we obtain an effective Lagrangian by integrating the PL density in the radial (axial) directions and write down the corresponding Euler-Lagrange equations (ELEs). We assume that the system is in the stationary regime and suppose to create a perturbation - with respect to the equilibrium - sufficiently weak so to retain in the ELEs only the first-order perturbation terms. We then numerically determine the behavior of the sound velocity $c_s$ as a function of the equilibrium density in the cigar-(disk)- shaped configuration when $y \ll -1$ (BCS limit) and when $y \gg 1$ (BEC limit). Note that on the BEC side, due to formation of two-body molecular bound states (see the above discussion), the binding energy  $\epsilon_{B}=\hbar^2/ma_{s}^2$ of each molecular dimer ($m$ being the mass of the fermions) will contribute to the chemical potential $\mu$. The energy $\epsilon_B$ does not depend on the space so that it will not affect the properties of the system since, as we shall see in the following, in the ELE corresponding to the equation of motion there are the spatial derivatives of $\mu$. 

For each of the two above cases, i.e. BCS limit and BEC limit, we investigate the role played by the dimensionality in determining the atomic cloud properties. Therefore we compare the above numerical solutions both with those pertaining to the three dimensional cases (corresponding to the very large densities limit) and pure 1D(2D) case (where the radial (axial) cloud width equal to the radial (axial) characteristic harmonic length). By this comparison, we determine the density-range validity of our numerical solutions by pointing out these latter solutions describe both the extreme density regimes and the intermediate regions.

\section{The system}
We study a superfluid gas of $N$ fermions confined by an external three-dimensional potential $U(\bm{r})$ at zero temperature. To do this we use the superfluid hydrodynamic Popov Lagrangian density \cite{popov}
\begin{equation}
\label{lagrangiandensity}
\mathcal{L}=-\hbar\dot{\theta}n-\frac{\hbar^{2}}{2m}(\bm{\nabla}\theta)^{2}n-U(\bm{r})n-\mathcal{E}(n,\bm{\nabla}n)n
\;,\end{equation}
where $m$ is the mass of the fermions and both the density $n$ and the phase $\theta$ depend on the position $\bm{r}$ and time $t$, that is $n=n(\bm{r},t)$ and $\theta=\theta(\bm{r},t)$. The quantity $\mathcal{E}(n,\bm{\nabla}n)$ represents the internal energy of the system that we write as a polytropic equation of the state plus a gradient correction which reproduces the effect of the quantum pressure, that is
\begin{equation}
\label{internal}
\mathcal{E}(n,\bm{\nabla}n)=\frac{\alpha}{\gamma}n^{\gamma-1}+\lambda\frac{\hbar^{2}}{8m}\frac{(\bm{\nabla} n)^2}{n^{2}}
\;.\end{equation}
The parameter $\gamma$, as shown in Ref. \cite{politropic}, varies within a specific range during the BCS-BEC crossover. $\gamma=5/3$ ($\displaystyle \alpha=\frac{3}{5}\frac{\hbar^{2}}{2m}(3\pi^{2})^{\frac{2}{3}}$) when $y=(k_Fa_s)^{-1} \ll -1$ (BCS limit), while to $\gamma=2$ ($\displaystyle \alpha=\frac{4\pi\hbar^{2}a_{s}N}{m}$ with $a_s$ the s-wave scattering length) when $y \gg 1$ (BEC limit). The intensity of the gradient correction is controlled by the parameter $\lambda$ \cite{weiz,kirzhnits,holas,salasnich3,salasnichnse1}.\\

In the following, we consider two confinement configurations: the cigar-shaped configuration and the disk-shaped one.

\section{Cigar-shaped configuration}
We assume that the trapping potential $U(\bm{r})$ is given by the superposition of an isotropic harmonic confinement in the radial ($x-y$) plane and a generic potential $V(z)$ in the axial ($z$) direction, so that
\begin{equation}
\label{cigar}
U(\bm{r})=\frac{1}{2}m\omega^{2}_{\perp}(x^{2}+y^{2})+V(z)
\;,\end{equation}
where $\omega_{\perp}$ is the trapping frequency in the $x-y$ plane. We make the hypothesis of a strong transverse harmonic confinement and a weak axial potential, so that it is possible to achieve a cigar-shaped configuration. Due to the form (\ref{cigar}) of the confining  potential, we can perform on the density and phase the following ansatz \cite{salasnich1}
\begin{eqnarray}
\label{cigaransatz}
&&n(\bm{r},t)=n_{0}(x,y,\sigma(z,t))n_{1}(z,t)=\frac{1}{\pi\sigma^{2}}e^{-\frac{x^{2}+y^{2}}{\sigma^{2}}}n_{1}(z,t)\nonumber\\
&&\theta(\bm{r},t)=\theta(z,t)
\;,\end{eqnarray}
where $\int dxdy n_0=1$ and $\int d \bm{r} n(\bm{r},t)=N$, and $\sigma$ represents the width of the gas cloud in the $x$ and and $y$ directions. By using the ansatz (\ref{cigaransatz}) at the right-hand side of Eq.(\ref{lagrangiandensity}) and integrating the resulting Lagrangian density in the transverse directions, i.e. $\int dxdy \mathcal{L}$, we get the effective Lagrangian
\begin{eqnarray}
\label{lcigar}
\tilde{\mathcal{L}}(n_{1},\theta,\sigma) &=& -n_{1}\left[\hbar\dot{\theta}+\frac{\hbar^{2}}{2m}\left(\frac{\partial\theta}{\partial z}\right)^{2}+\frac{1}{2}m\omega^{2}_{\perp}\sigma^{2}+V(z)\right]\nonumber\\
&-&\frac{\alpha n_{1}^{\gamma}}{\gamma^{2}\pi^{\gamma-1}\sigma^{2(\gamma-1)}}
-\frac{\lambda\hbar^{2}}{8mn_{1}}\left(\frac{\partial n_{1}}{\partial z}\right)^{2}\nonumber\\
&-&\frac{\lambda\hbar^{2}n_{1}}{2m\sigma^{2}}\left[1+\left(\frac{\partial\sigma}{\partial z}\right)^{2}\right]
\;.\end{eqnarray}

From this Lagrangian we derive the Euler-Lagrange equations (ELEs). The ELE with respect to $\theta$ is
\begin{equation}
\label{elethetacigar}
\frac{\partial \tilde{\mathcal{L}}}{\partial \theta}-\frac{\partial}{\partial t}\frac{\partial \tilde{\mathcal{L}}}{\partial \dot \theta}-
\frac{\partial}{\partial z}\frac{\partial \tilde{\mathcal{L}}}{\partial \partial_z \theta}=0
\;,\end{equation}
which gives rise to
\begin{equation}
\label{cecigar}
\frac{\partial n_{1}}{\partial t}+\frac{\partial}{\partial z}(n_{1}v_{z})=0
\;.\end{equation}
This is the continuity equation, where $v_{z}=\displaystyle{\frac{\hbar}{m}\frac{\partial\theta}{\partial z}}$.
The ELE with respect to $n_1$ is
\begin{equation}
\label{elen1cigar}
\frac{\partial \tilde{\mathcal{L}}}{\partial n_1}-\frac{\partial}{\partial t}\frac{\partial \tilde{\mathcal{L}}}{\partial \dot n_1}-
\frac{\partial}{\partial z}\frac{\partial \tilde{\mathcal{L}}}{\partial \partial_z n_1}=0
\;,\end{equation}
which, by neglecting the spatial derivatives of $\sigma$, provides the equation of motion
\begin{eqnarray}
\label{mecigar}
&&m\frac{\partial v_{z}}{\partial t}+\frac{\partial}{\partial z}\left[\frac{1}{2}mv_{z}^{2}+V(z)+\mu(n)\right]-\frac{\lambda\hbar^{2}}{4mn_{1}}\frac{\partial^{3}n_{1}}{\partial z^3}\nonumber\\
&+&\frac{\partial n_{1}}{\partial z}\left[\frac{\lambda\hbar^{2}}{2mn_{1}^{2}}\frac{\partial^{2}n_{1}}{\partial z^{2}}-\frac{\lambda\hbar^{2}}{4mn_{1}^{3}}\left(\frac{\partial n_{1}}{\partial z}\right)^{2}\right]=0
\;,\end{eqnarray}
where
\begin{equation}
\label{mucigar}
\mu(n)=\frac{\alpha}{\gamma}\left(\frac{n_{1}}{\pi\sigma^{2}}\right)^{\gamma-1}
\end{equation}
is the chemical potential.

Finally, let us focus on the role played by $\sigma$. This quantity can be viewed as a function of $z$ and $t$ and as a variational parameter as well. In the first case, i.e.
$\sigma=\sigma(z,t)$, we can start from the ELE with respect to $\sigma$
\begin{equation}
\label{elesigmacigar}
\frac{\partial\tilde{\mathcal{L}}}{\partial \sigma}-\frac{\partial}{\partial t}\frac{\partial \tilde{\mathcal{L}}}{\partial \dot \sigma}-
\frac{\partial}{\partial z}\frac{\partial \tilde{\mathcal{L}}}{\partial \partial_z \sigma}=0
\;.\end{equation}
One, at this point, performs a sort of adiabatic approximation - i.e. one neglects the derivatives of $\sigma$ with respect to space and time - from which one gets the following simple algebraic equation for $\sigma$:
\begin{equation}
\label{sigmacigar1}
\sigma^{2\gamma}-\lambda a^{4}_{\perp}\sigma^{2\gamma-4}-\frac{2\alpha(\gamma-1)n_{1}^{\gamma-1}}{\gamma^{2}\pi^{\gamma-1}m\omega^{2}_{\perp}}=0
\;,\end{equation}
where $a_{\perp}=\displaystyle{\sqrt{\frac{\hbar}{m\omega_{\perp}}}}$ is the transverse characteristic harmonic length.

In the second case, i.e. when $\sigma$ plays the role of a variational parameter, we have to minimize the action $\displaystyle S=\int dzdt\tilde{\mathcal{L}}$ with respect to $\sigma$. By requiring therefore that $\displaystyle{\frac{\partial S}{\partial \sigma}}=0$ we get the following equation:
\begin{equation}
\label{sigmacigar2}
\sigma^{2\gamma}-\lambda a^{4}_{\perp}\sigma^{2\gamma-4}-\frac{2\alpha(\gamma-1)}{\gamma^{2}\pi^{\gamma-1}m\omega^{2}_{\perp}}\frac{I_{\gamma}}{I_{1}}=0
\;,\end{equation}
where $\displaystyle I_{\delta}=\int dzdtn_{1}^{\delta}$. The Eqs. (\ref{sigmacigar1}) and (\ref{sigmacigar2}) provide a constraint to which $\sigma(n)$ has to obey. In conclusion, we observe that in Eq. (\ref{sigmacigar1}) appears the dependence on the space $z$ and time $t$, while this is not the case of Eq. (\ref{sigmacigar2}) because in $I_{\delta}$ (see above), $z$ and $t$ will be integrated out.

\section{Disk-shaped configuration}
For this case we suppose that the confining potential  $U(\bm{r})$ derives from the superposition of a harmonic potential in the axial direction $z$ and a generic potential, say $W(x,y)$, in the radial transverse plane ($x-y$). Then
\begin{equation}
\label{disk}
U(\bm{r})=\frac{1}{2}m\omega_{z}^{2}z^{2}+W(x,y)
\;,\end{equation}
where $\omega_z$ is the trapping frequency in the $z$ direction. We suppose a strong confinement in the axial direction, while the potential in the radial plane is assumed to be weak, so that we have to a disk-shaped configuration. The form (\ref{disk}) of the trapping potential allows us to perform on the density and the phase the following ansatz
\begin{eqnarray}
\label{diskansatz}
&&n(\bm{r},t)=n_{0}(z,\eta(x,y,t))n_{1}(x,y,t)=\frac{1}{\sqrt{\pi}\eta}e^{-\frac{z^{2}}{\eta^{2}}}n_{1}(x,y,t) \nonumber\\
&&\theta(\bm{r},t)=\theta(x,y,t)
\;,\end{eqnarray}
where $\int dz n_0=1$ and $\int d\bm{r} n(\bm{r},t)=N$, and $\eta$ represents the width of the gas cloud in the $z$ direction. By employing the ansatz (\ref{diskansatz}) at the right-hand side of Eq. (\ref{lagrangiandensity}) and performing the integration of the so obtained Lagrangian density in the axial direction, i.e. $\int dz \mathcal{L}$, we get the following effective Lagrangian
\begin{eqnarray}
\label{ldisk}
\tilde{\mathcal{L}}(n_{1},\theta,\eta) & = & -n_{1}\left[\hbar\dot{\theta}+\frac{\hbar^{2}}{2m}(\bm{\nabla}_{\bm{R}} \,\theta)^{2}+\frac{1}{4}m\omega^{2}_{z}\eta^{2}+W(x,y)\right]
\nonumber\\
&-&\frac{\alpha n_{1}^{\gamma}}{\gamma^{\frac{3}{2}}\pi^{\frac{\gamma-1}{2}}\eta^{\gamma-1}}
-\frac{\lambda\hbar^{2}}{8mn_{1}}(\bm{\nabla}_{\bm{R}} n_{1})^{2}-\frac{\lambda\hbar^{2}n_{1}}{4m\eta^{2}}\left[1+(\bm{\nabla}_{\bm{R}}\,\eta)^{2}\right]\nonumber\\
\end{eqnarray}
with $\bm{R}=(x,y)$ and $\bm{\nabla}_{\bm{R}}=\displaystyle{(\frac{\partial}{\partial x},\frac{\partial}{\partial y})}$.
From the Lagrangian (\ref{ldisk}) one derives the corresponding Euler-Lagrange equations. By following the same procedure followed in the previous section, the continuity equation is achieved from the ELE with respect to $\theta$
\begin{equation}
\label{elethetadisk}
\frac{\partial \tilde{\mathcal{L}}}{\partial \theta}-\frac{\partial}{\partial t}\frac{\partial \tilde{\mathcal{L}}}{\partial \dot \theta}-
\bm{\nabla}_{\bm{R}}\cdot\frac{\partial \tilde{\mathcal{L}}}{\partial \bm{\nabla}_{\bm{R}}\,\theta}=0
\;\end{equation}
which becomes
\begin{equation}
\label{cedisk}
\frac{\partial n_{1}}{\partial t}+\bm{\nabla}_{\bm{R}}\,\cdot(n_{1}\bm{v})=0
\;,\end{equation}
where $\bm{v}=\displaystyle{\frac{\hbar}{m}\bm{\nabla}_{\bm{R}}\,\theta}$.

The motion equation is provided from the Euler-Lagrange equation with respect to $n_1$:
\begin{equation}
\label{elen1disk}
\frac{\partial \tilde{\mathcal{L}}}{\partial n_1}-\frac{\partial}{\partial t}\frac{\partial \tilde{\mathcal{L}}}{\partial \dot n_1}-
 \bm{\nabla}_{\bm{R}}\,\cdot\frac{\partial \tilde{\mathcal{L}}}{\partial  \bm{\nabla}_{\bm{R}}\, n_1}=0
\;,\end{equation}
which, by neglecting the spatial derivatives of $\eta$, reads
\begin{eqnarray}
\label{medisk}
&&m\frac{\partial\bm{v}}{\partial t}+\bm{\nabla}_{\bm{R}}\,\left[\frac{m\bm{v}^2}{2}+W(x,y)+\mu(n)\right]\nonumber\\
&-&\frac{\lambda\hbar^2}{4mn_{1}}\bm{\nabla}_{\bm{R}}\,(\bm{\nabla}_{\bm{R}}^{2}n_{1})\nonumber\\
&+&\bm{\nabla}_{\bm{R}}\,n_{1}\left[\frac{\lambda\hbar^2}{2mn_{1}^{2}}\bm{\nabla}_{\bm{R}}^{2}\,n_{1}-\frac{\lambda\hbar^2}{4mn_{1}^{3}}(\bm{\nabla}_{\bm{R}}\,n_{1})^{2}\right]=0
\;\end{eqnarray}
with
\begin{equation}
\label{mudisk}
\mu(n)=\frac{\alpha}{\sqrt{\gamma}}\left(\frac{n_{1}}{\sqrt{\pi}\eta}\right)^{\gamma-1}
\;\end{equation}
the chemical potential.

Finally, $\eta$. As in the cigar-shaped case, $\eta$ can be interpreted as a function of the space and time, i.e $\eta=\eta(x,y,t)$, and also as a variational parameter. In the first case, we start from the Euler-Lagrange equation with respect to $\eta$
\begin{equation}
\label{elesigmadisk}
\frac{\partial \tilde{\mathcal{L}}}{\partial \eta}-\frac{\partial}{\partial t}\frac{\partial \tilde{\mathcal{L}}}{\partial \dot \eta}-
\bm{\nabla}_{\bm{R}}\,\cdot \frac{\partial \tilde{\mathcal{L}}}{\partial \bm{\nabla}_{\bm{R}}\, \eta}=0
\;.\end{equation}
This equation, by neglecting the derivatives of $\eta$ with respect to space and time, becomes
\begin{equation}
\label{etadisk1}
\eta^{\gamma+1}-\lambda a^{4}_{z}\eta^{\gamma-3}-\frac{2\alpha(\gamma-1)n_{1}^{\gamma-1}}{\gamma^{\frac{3}{2}}\pi^{\frac{\gamma-1}{2}}m\omega^{2}_{z}}=0
\;,\end{equation}
where $a_{z}=\displaystyle{\sqrt{\frac{\hbar}{m\omega_{z}}}}$ is the axial harmonic characteristic length.

If $\eta$ is, instead, interpreted as a variational parameter, then we minimize the action with respect to $\eta$ and get
\begin{equation}
\label{etadisk2}
\eta^{\gamma+1}-\lambda a^{4}_{z}\eta^{\gamma-3}-\frac{2\alpha(\gamma-1)}{\gamma^{\frac{3}{2}}\pi^{\frac{\gamma-1}{2}}m\omega^{2}_{z}}\frac{I_{\gamma}}{I_{1}}=0
\;\end{equation}
with $\displaystyle I_{\delta}=\int dxdydtn_{1}^{\delta}$. Due to this latter integral, the space-time dependence does not appear in Eq. (\ref{etadisk2}).

\section{Atomic cloud properties}
To gain a deeper physical insight about our system, we focus on the study of the sound velocity $c_s$. This latter is strictly related to the propagation of a perturbation, with respect to the equilibrium, created in a given point of the system at a given time. We analyze $c_s$ as a function of the spatial density both in the cigar-shaped configuration and in the disk-shaped one. As first, we derive the sound velocity in a general scenario, that is starting from a general external potential $U(\bm{r})$.

\subsection{Sound velocity: general treatment}
We start from the the Lagrangian density (\ref{lagrangiandensity}). Through the related Euler-Lagrangian equations we achieves two equations for the spatial density $n(\bm{r},t)$ and the velocity field $\bm{v}(\bm{r},t)$:
\begin{equation}
\label{1stg}
\frac{\partial n}{\partial t}+\bm{\nabla}\cdot(n\bm{v})=0
\;,\end{equation}
and
\begin{equation}
\label{2ndg}
m \frac{\partial \bm{v}}{\partial t}+ \bm{\nabla}\bigg[\frac{1}{2}m \bm{v}^2+ U(\bm{r})+X(n,\bm{\nabla}n)\bigg]=0
\;,\end{equation}
where $\displaystyle{X(n,\bm{\nabla} n)=\alpha n^{\gamma-1}-\lambda \frac{\hbar^2}{2m}\frac{\nabla^2 \sqrt{n}}{\sqrt{n}}}$. We assume to be in the stationary regime, i.e. $\bm{v}=\bm{0}$, and in the uniform case, i.e. $U(\bm{r})=0$. At this point, let us suppose to perturbate the system with respect to the equilibrium configuration characterized by
$n(\bm{r},t)=n_{eq}$ and $\bm{v}=\bm{0}$, that is
\begin{eqnarray}
\label{perturbation}
&&n(\bm{r},t)=n_{eq}+ \tilde n(\bm{r},t)\nonumber\\
&& \bm{v}(\bm{r},t)=\bm{0}+\bm{\tilde v} (\bm{r},t)
\;.\end{eqnarray}
We use these equations in the ELEs (\ref{1stg}) and (\ref{2ndg}). Under the hypothesis that the perturbation is sufficiently weak so to retain only the first-order perturbation terms in the ELEs, we get
\begin{equation}
\label{1stg2}
\frac{\partial \tilde n}{\partial t}+n_{eq}\bm{\nabla}\cdot\bm{\tilde v}=0
\;,\end{equation}
and
\begin{equation}
\label{2ndg2}
n_{eq} \frac{\partial \bm{\tilde v} }{\partial t}+\frac{\alpha(\gamma-1)n_{eq}^{\gamma-1}}{m} \bm{\nabla} \tilde n-\lambda\frac{\hbar^2}{4m^2} \bm{\nabla} \nabla^2 \tilde n=0
\;.\end{equation}
By calculating the time derivative of Eq. (\ref{1stg2}) and the divergence of Eq. (\ref{2ndg2}) and subtracting the former equation from the latter, we get
\begin{equation}
\label{final}
\bigg[\frac{\partial^2}{\partial t^2} -\frac{\alpha(\gamma-1)n_{eq}^{\gamma-1}}{m} \nabla^2+\lambda \frac{\hbar^2}{4m^2}\nabla^2 \nabla^2\bigg]\tilde n=0
\;.\end{equation}
This equation describes the spatio-temporal propagation of the wave associated to the perturbation. The propagation of the wave pressure takes place with velocity $c_s$ (sound velocity) which is given by the squared root of the coefficient of the Laplacian in Eq. (\ref{final}) (that is also the coefficient of $\bm{\nabla} \tilde n$ in Eq. (\ref{2ndg2})), i.e. 
\begin{equation}
\label{sound3d2}
c_{s}=\big(\frac{\alpha(\gamma-1)n_{eq}^{\gamma-1}}{m}\big)^{\frac{1}{2}}
\;.\end{equation}
Note that Eq. (\ref{final}) is different from the well-known equation for the spatio-temporal propagation of the waves. In fact, this equation contains an additional term, i.e. that controlled by the parameter $\lambda$.

At this point, it is worth to observe that if the perturbation is a plane wave, that is $\tilde n(\bm{r},t)=Ae^{i(\bm{k}\cdot \bm{r}-\omega t)}+A^{*}e^{-i(\bm{k}\cdot \bm{r}-\omega t)}$, the relation of dispersion which characterizes the oscillations associated to the sound wave induced by the perturbation is
\begin{equation}
\label{dispersion}
\omega=c_sk\sqrt{1+\frac{\lambda\hbar^2}{4m^2c_{s}^{2}}k^2}
\;,\end{equation}
where $c_s$ is the sound velocity (\ref{sound3d2}). In the limit of sufficiently small wave vector ($k\rightarrow 0$), Eq. (\ref{dispersion}) gives back the usual dispersion relation of the sound wave in the absence of the gradient correction.

\subsection{Sound velocity in the cigar-shaped configuration}
Let us start from the external trapping potential (\ref{cigar}). We assume that $V(z)=0$ and to be in the stationary regime, i.e. $v=0$. We suppose to perturbate the system with respect to the equilibrium configuration where $n_1(z,t)=n_{eq}$ and $v(z,t)=0$:
\begin{eqnarray}
\label{perturbationcigar}
&&n_1(z,t)=n_{eq}+ \tilde n(z,t)\nonumber\\
&& v_z(z,t)=0+\tilde v (z,t)
\;.\end{eqnarray}
Then, we use the two above in equations in Eq. (\ref{cecigar}) (continuity equation) and in Eq. (\ref{mecigar}) (motion equation) that after the linearization (we are supposing that the perturbation is sufficiently weak so that to retain only the first-order perturbation terms) provide
\begin{equation}
\label{cecigars}
\frac{\partial \tilde n}{\partial t}+n_{eq}\frac{\partial \tilde v}{\partial z}=0
\;,\end{equation}
and
\begin{equation}
\label{mecigars}
n_{eq} \frac{\partial \tilde v }{\partial t}+\frac{\alpha(\gamma-1)n_{eq}^{\gamma-1}}{m \gamma\pi^{\gamma-1}\sigma^{2(\gamma-1)}} \frac{\partial \tilde n}{\partial z}-\lambda\frac{\hbar^2}{4m^2} \frac{\partial^3 \tilde n}{\partial z^3} =0
\;.\end{equation}
Then (see the discussion of the previous subsection after Eqs. (\ref{2ndg2}) and (\ref{final})) the sound velocity $c_s$ is
\begin{equation}
\label{soundcigar}
c_{s}=\bigg(\frac{\alpha(\gamma-1)}{m\gamma}\bigg)^{\frac{1}{2}}\left(\frac{n_{eq}}{\pi\sigma^2(n_{eq})}\right)^{(\gamma-1)/2}
\;,\end{equation}
where
the radial width $\sigma$ is constrained according Eq. (\ref{sigmacigar1}) which (with $n_1$ replaced by $n_{eq}$) reads
\begin{equation}
\label{sigmaconst}
\sigma^{2\gamma}-\lambda a^{4}_{\perp}\sigma^{2\gamma-4}-\frac{2\alpha(\gamma-1)n_{eq}^{\gamma-1}}{\gamma^{2}\pi^{\gamma-1}m\omega^{2}_{\perp}}=0
\;.\end{equation}

\subsection{Sound velocity in disk-shaped configuration}
In this case the trapping potential is given by Eq. (\ref{disk}). We assume that is $W(x,y)=0$ and to be in the stationary regime: $\bm{v}(x,y,t)=\bm{0}$. As before, let us suppose to create a small perturbation in the system with respect to the equilibrium,
\begin{eqnarray}
\label{perturbationdisk}
&&n(x,y,t)=n_{eq}+ \tilde n(x,y,t)\nonumber\\
&& \bm{v}(x,y,t)=\bm{0}+\bm{\tilde v} (x,y,t)
\;.\end{eqnarray}
By using these two equations in the continuity equation (\ref{cedisk}) and in he motion equation (\ref{medisk}) and neglecting the quadratic terms in the perturbation, we get
\begin{equation}
\label{cedisks}
\frac{\partial \tilde n}{\partial t}+n_{eq}\bm{\nabla}\cdot\bm{\tilde v}=0
\;,\end{equation}
and
\begin{equation}
\label{medisk2}
n_{eq} \frac{\partial \bm{\tilde v} }{\partial t}+\frac{\alpha(\gamma-1)n_{eq}^{\gamma-1}}{m \gamma^{1/2}\pi^{(\gamma-1)/2}\eta^{\gamma-1)}} \bm{\nabla} \tilde n-\lambda\frac{\hbar^2}{4m^2} \bm{\nabla} \nabla^2 \tilde n=0
\;.\end{equation}
In this case, by employing the same arguments followed in the two previous subsections, the sound velocity $c_s$ is
\begin{equation}
\label{sounddisk}
c_{s}=\bigg(\frac{\alpha(\gamma-1)}{m\gamma^{\frac{1}{2}}}\bigg)^{\frac{1}{2}}\left(\frac{n_{eq}}{\pi^{\frac{1}{2}}\eta(n_{eq})}\right)^{(\gamma-1)/2}
\;\end{equation}
with
the axial width $\eta$ has to satisfy Eq. (\ref{etadisk1}), that, for this case (with $n_1$ replaced by $n_{eq}$), has the form
\begin{equation}
\label{etaconst}
\eta^{\gamma+1}-\lambda a^{4}_{z}\eta^{\gamma-3}-\frac{2\alpha(\gamma-1)n_{eq}^{\gamma-1}}{\gamma^{\frac{3}{2}}\pi^{\frac{\gamma-1}{2}}m\omega^{2}_{z}}=0
\;.\end{equation}

\section{Analysis}
To analyze the behavior of the sound velocity as a function of the density, one has to find $\sigma(n_{eq})$ and $\eta(n_{eq})$  - from the constraints (\ref{sigmaconst}) and (\ref{etaconst}) - that employed at the right-hand side of Eqs. (\ref{soundcigar}) and (\ref{sounddisk}) will provide $c_s=c_s(n_{eq})$.
We study the widths of the gas cloud and the sound velocity both in the cigar-shaped configuration and in the  disk-shaped one in two cases by assuming that the intensity of gradient correction $\lambda$ is equal to $1$.

The first case that we analyze is $\gamma=\displaystyle{\frac{5}{3}}$ ($ y \ll -1$, BCS limit) when $\displaystyle \alpha=\frac{3}{5}\frac{\hbar^{2}}{2m}(3\pi^{2})^{\frac{2}{3}}$. In this situation, for the cigar-shaped configuration, Eqs. (\ref{soundcigar}) and (\ref{sigmaconst}) become
\begin{equation}
\label{3d1dc1}
c_s=\frac{(3)^{1/2} (9\pi^2)^{1/6}}{5}\frac{\hbar}{m}\big(\frac{n_{eq}}{\sigma^2}\big)^{\frac{1}{3}}
\;,\end{equation}
and
\begin{equation}
\label{3d1dw1}
\sigma^4-\frac{18 (9\pi^2)^{1/3}}{125}a^{4}_{\bot}n^{\frac{2}{3}}_{eq}\sigma^{\frac{2}{3}}-a^{4}_{\bot}=0
\;.\end{equation}
In the disk-shaped configuration, Eqs. (\ref{sounddisk}) and (\ref{etaconst}) read
\begin{equation}
\label{3d2dc1}
c_s=\frac{\pi^{1/2}3^{7/12}}{5^{3/4}}\frac{\hbar}{m}\big(\frac{n_{eq}}{\eta}\big)^{\frac{1}{3}}
\;,\end{equation}
and
\begin{equation}
\label{3d2dw1}
\eta^4-\frac{18\pi 3^{1/6}}{5^{5/2}}a^{4}_{z}n^{\frac{2}{3}}_{eq}\eta^{\frac{4}{3}}-a^{4}_{z}=0
\;.\end{equation}
\begin{figure}
\centering
\subfloat
	{\includegraphics[scale=0.6]{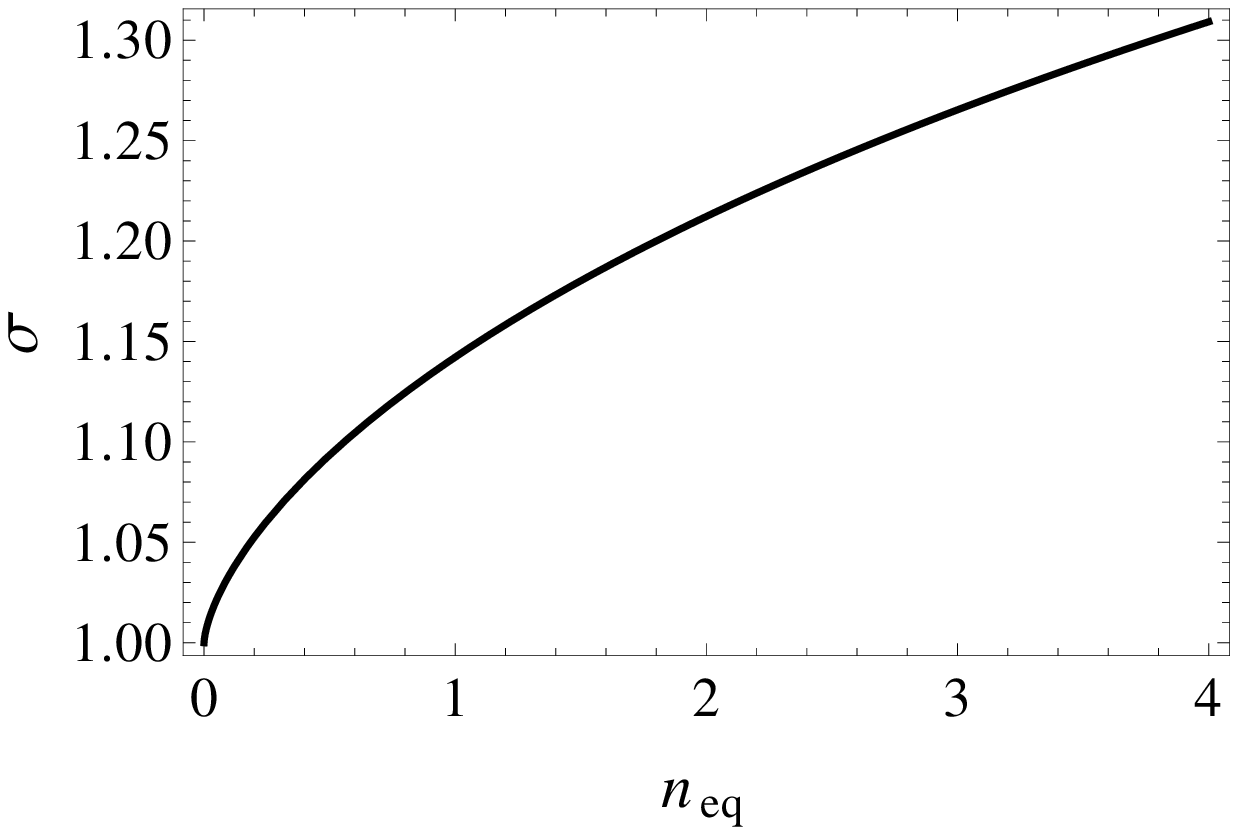}}
\subfloat
	{\includegraphics[scale=0.6]{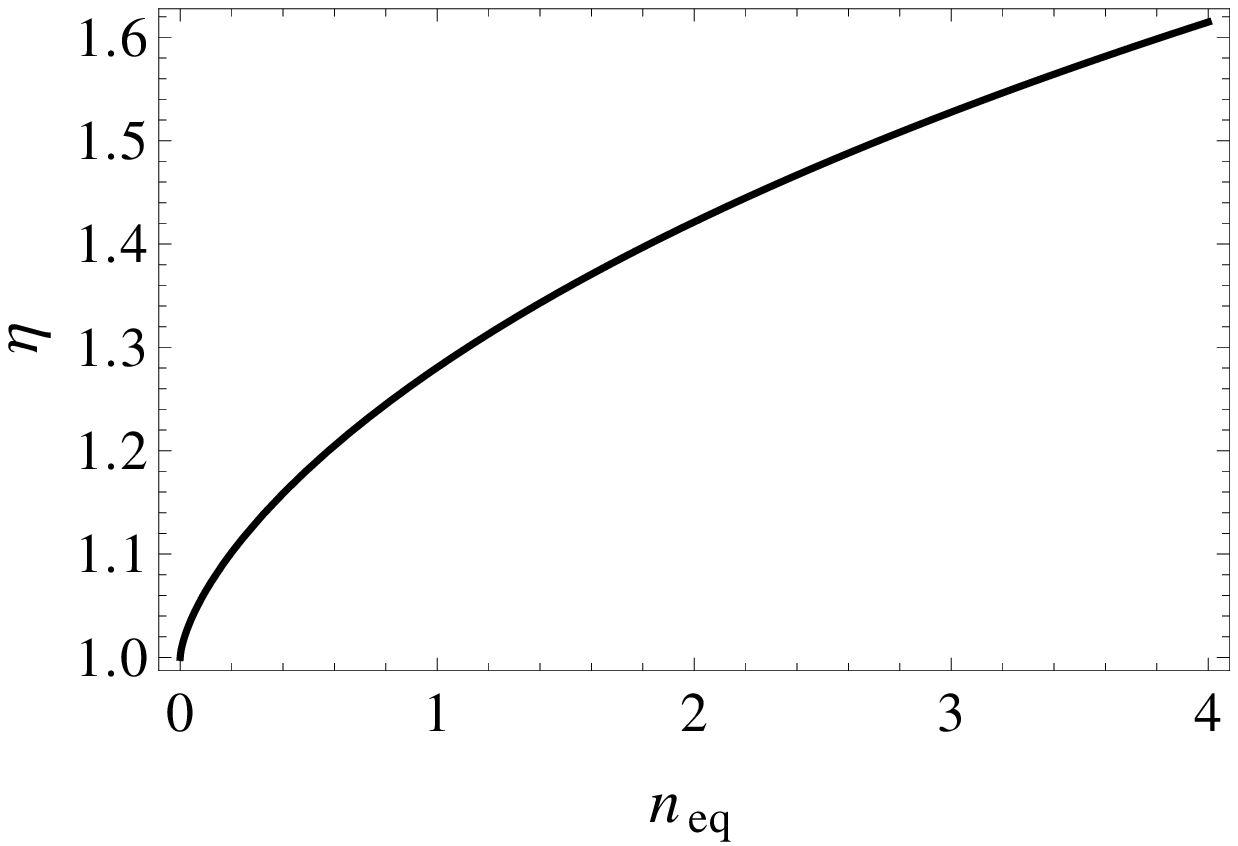}} \\
\subfloat
	{\includegraphics[scale=0.6]{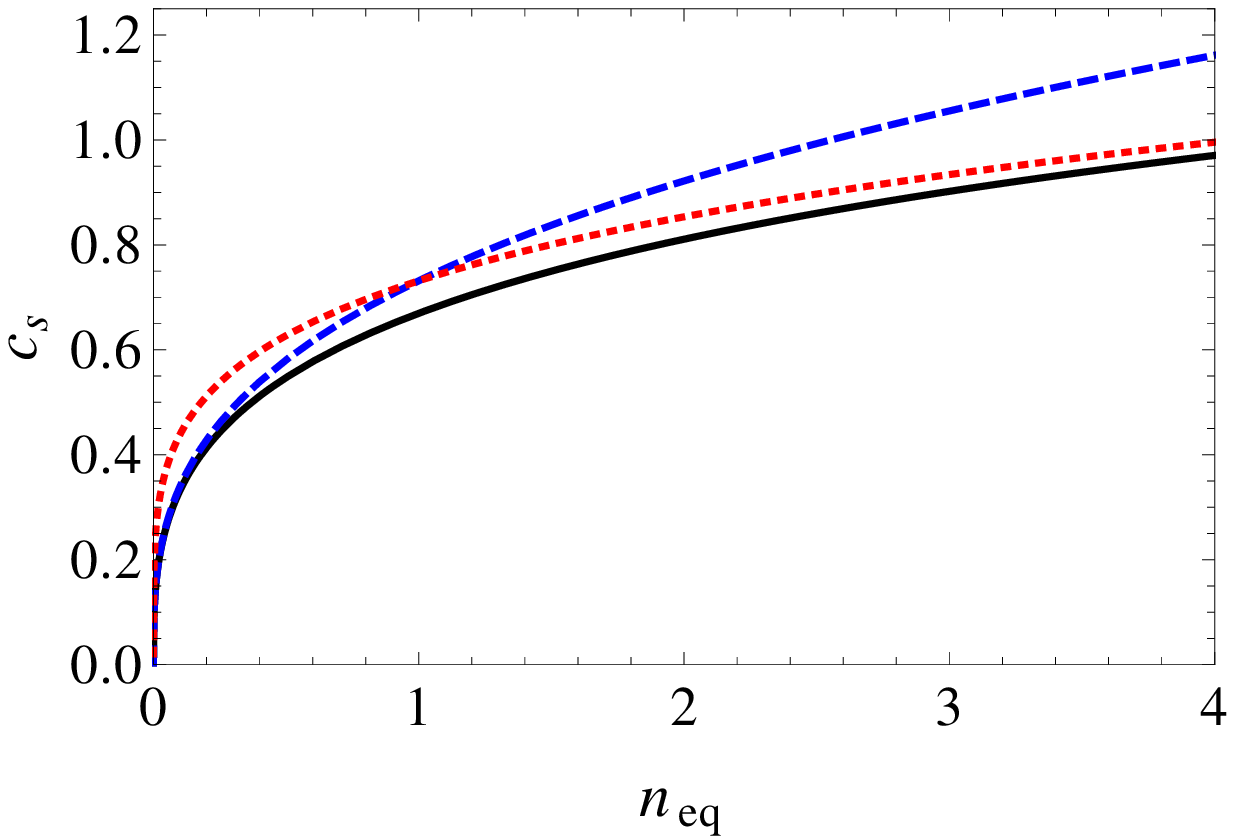}}
\subfloat
	{\includegraphics[scale=0.6]{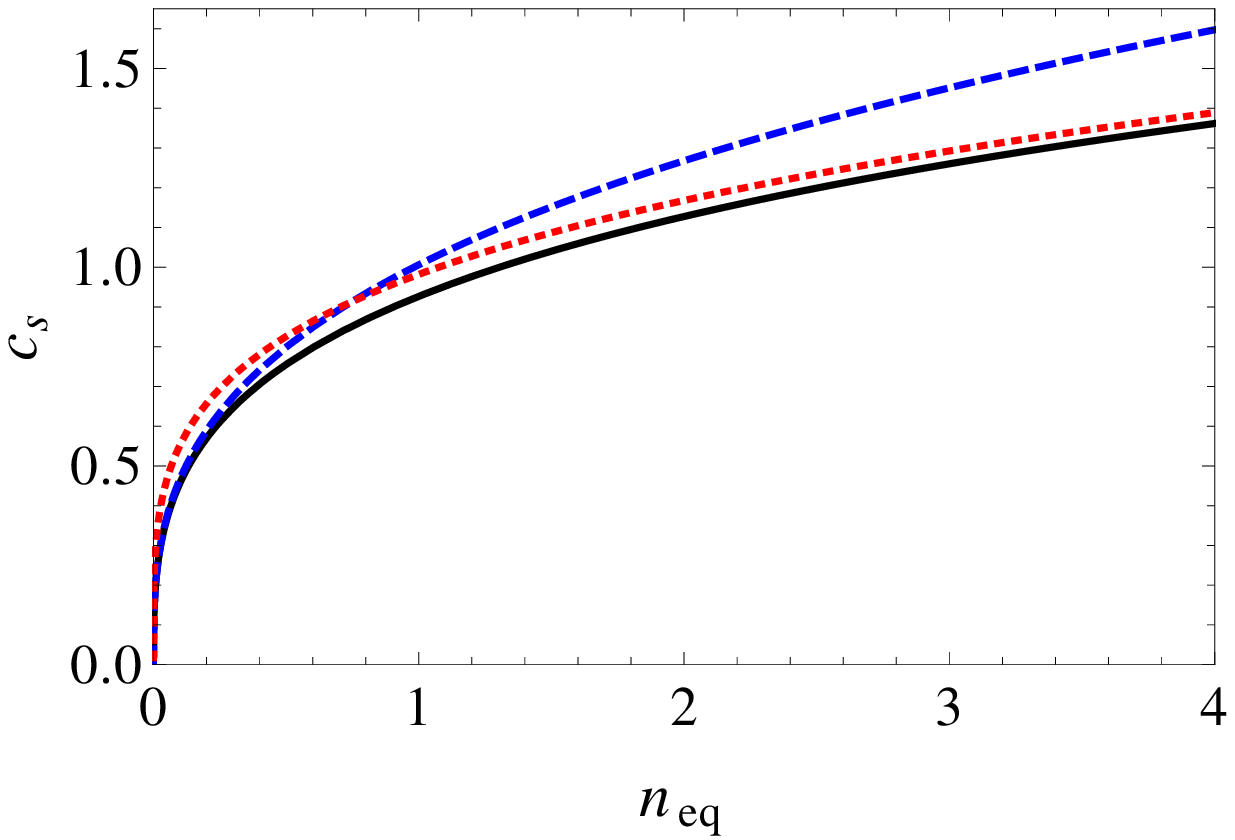}}
\caption{\small{Cigar-shaped (CS) case and disk-shaped (DS) one with $\gamma=5/3$ and $\lambda=1$. Left: CS configuration. Top panel: radial width $\sigma$ vs $n_{eq}$, Eq. (\ref{3d1dw1}). Bottom panel: sound velocity $c_s$ vs $n_{eq}$. Solid line: Eq. (\ref{3d1dc1}) solved with the constraint (\ref{3d1dw1}). Dashed line: 1D regime, Eq. (\ref{3d1dc1}) with $\sigma=1$. Dotted line: 3D regime (see the text). $\sigma$ is in units of $a_{\bot}$, $n_{eq}$ in units of $a^{-1}_{\bot}$, $c_s$ in units of $\omega_{\bot}a_{\bot}$. Right: DS configuration. Top panel: axial width $\eta$ vs $n_{eq}$, Eq. (\ref{3d2dw1}). Bottom panel: $c_s$ vs $n_{eq}$. Solid line: Eq. (\ref{3d2dc1}) solved with the constraint (\ref{3d2dw1}). Dashed line: 2D regime, Eq. (\ref{3d2dc1}) with $\eta=1$. Dotted line: 3D regime (see the text). $\eta$ is in units of $a_{z}$, $n_{eq}$ in units of $a^{-1}_{z}$, $c_s$ in units of $\omega_{z}a_{z}$.}}
\end{figure}

\begin{figure}
\centering
\subfloat
	{\includegraphics[scale=0.6]{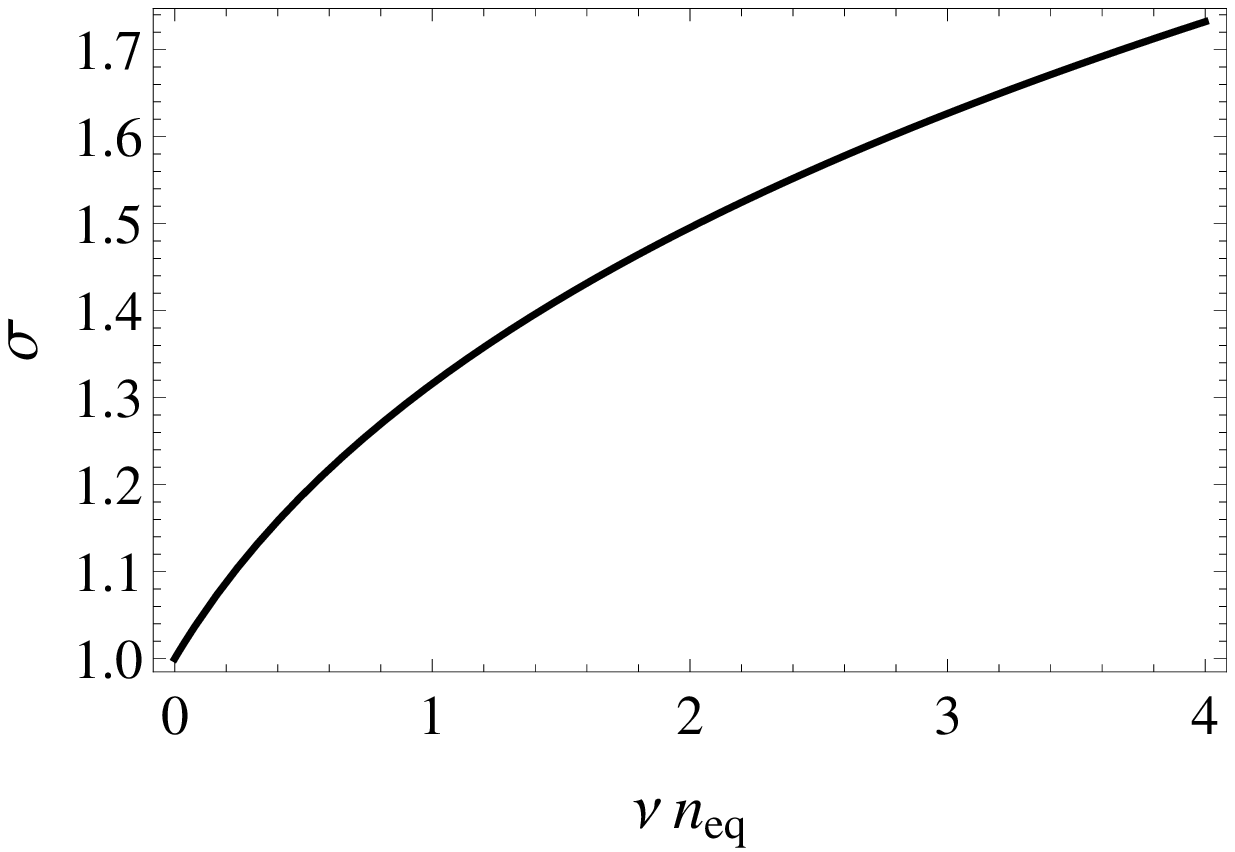}}
\subfloat
	{\includegraphics[scale=0.6]{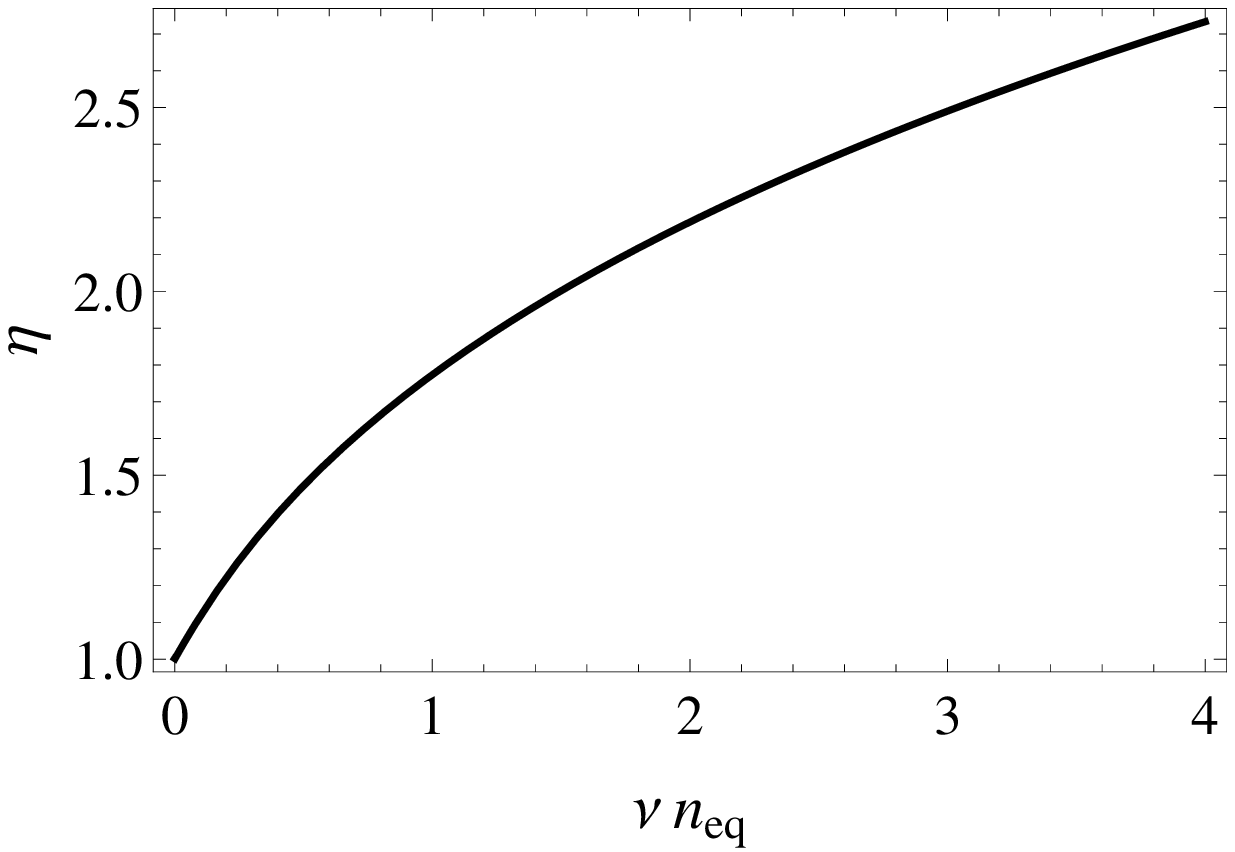}} \\
\subfloat
	{\includegraphics[scale=0.6]{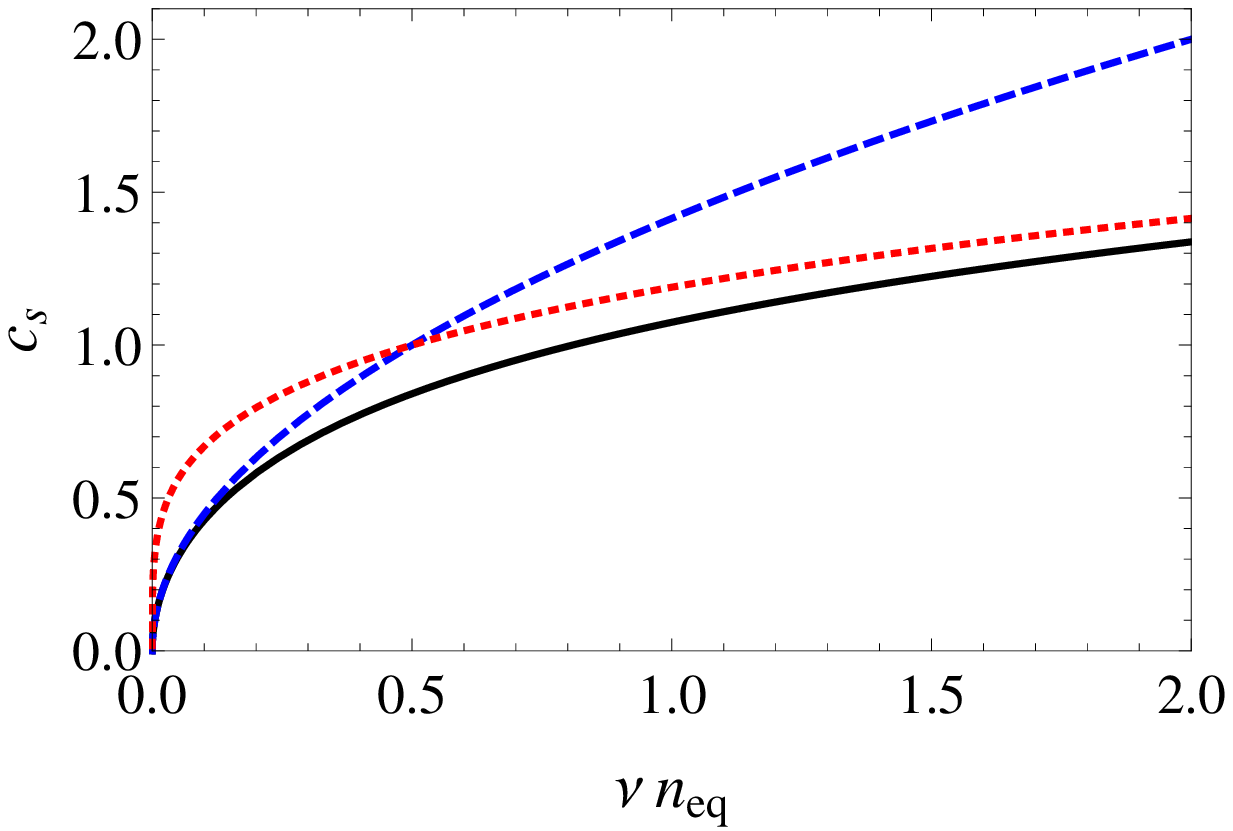}}
\subfloat
	{\includegraphics[scale=0.6]{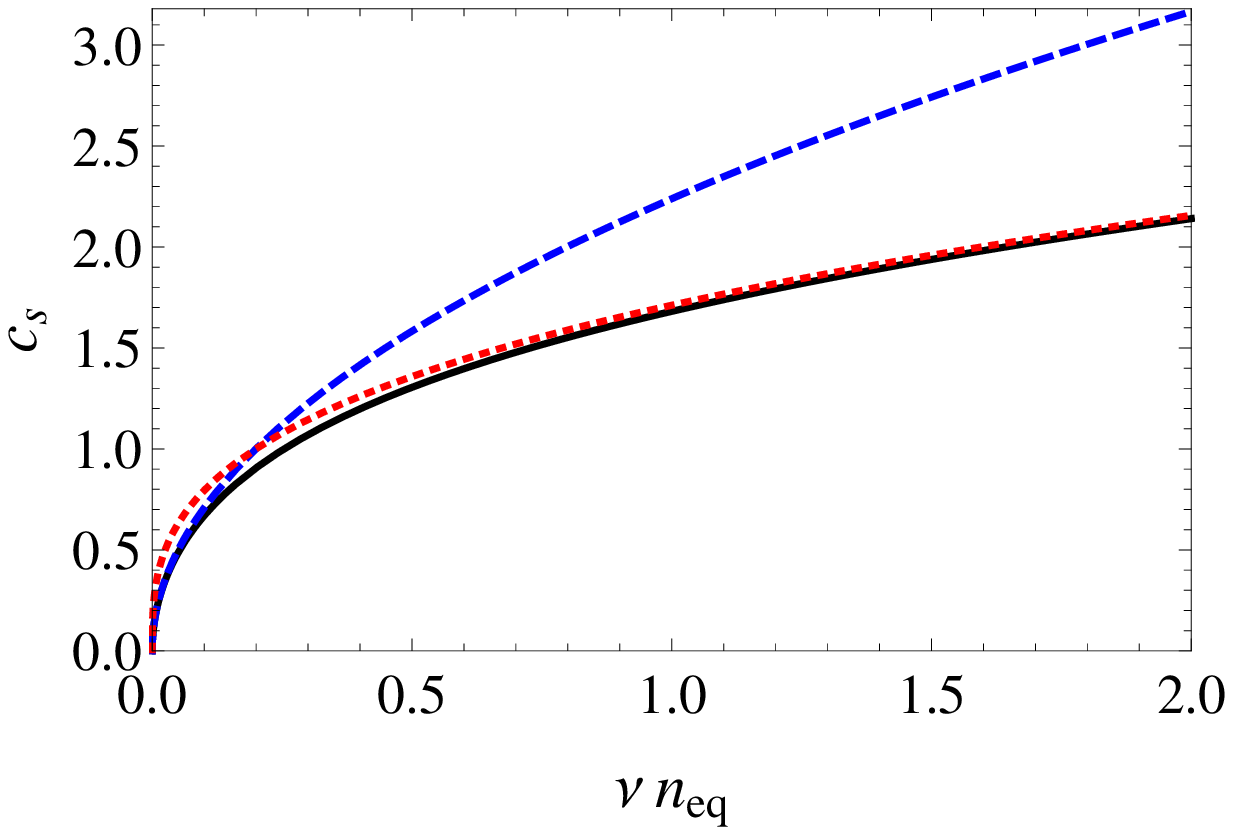}}
\caption{\small{Cigar-shaped (CS) case and disk-shaped (DS) one with $\gamma=2$ and $\lambda=1$. Left: CS configuration. Top panel: radial width $\sigma$ vs $\nu n_{eq}$ 
($\nu=a_sN$, $a_s$ in units of $a_{\bot}$), Eq. (\ref{3d1dw2}). Bottom panel: sound velocity $c_s$ vs $\nu n_{eq}$. Solid line: Eq. (\ref{3d1dc2}) solved with the constraint (\ref{3d1dw2}). Dashed line: 1D regime, Eq. (\ref{3d1dc2}) with $\sigma=1$. Dotted line: 3D regime (see the text). Units as in Fig.1. Right: DS configuration. Top panel: axial width $\eta$ vs $\nu n_{eq}$ ($\nu=a_sN$, $a_s$ in units of $a_{z}$), Eq. (\ref{3d2dw2}). Bottom panel: sound velocity $c_s$ vs $\nu n_{eq}$. Solid line: Eq. (\ref{3d2dc2}) solved with the constraint (\ref{3d2dw2}). Dashed line: 2D regime, Eq. (\ref{3d2dc2}) with $\eta=1$. Dotted line: 3D regime (see the text). Units as in Fig.1}}
\end{figure}

The second case that we deal with is $\gamma=2$ ($y \gg 1$, BEC limit) when $\displaystyle \alpha=\frac{4\pi\hbar^{2}a_{s}N}{m}$. Then, for the cigar-shaped configuration Eqs. (\ref{soundcigar}) and (\ref{sigmaconst}) will read
\begin{equation}
\label{3d1dc2}
c_s=(2)^{1/2}(a_s N)^{1/2}\frac{\hbar}{m}\big(\frac{n_{eq}}{\sigma^2}\big)^{\frac{1}{2}}
\;,\end{equation}
\begin{equation}
\label{3d1dw2}
\sigma^4-a^{4}_{\bot}(1+2a_sNn_{eq})=0
\;.\end{equation}
In the disk-shaped configuration, Eqs. (\ref{sounddisk}) and (\ref{etaconst}) assume the following form:
\begin{equation}
\label{3d2dc2}
c_s=2^{3/4} \pi^{1/4} \frac{\hbar}{m} (a_sN)^{1/2} \big(\frac{n_{eq}}{\eta}\big)^{\frac{1}{2}}
\;,\end{equation}
\begin{equation}
\label{3d2dw2}
\eta^{4}-2(2\pi)^{1/2}a^4_{z}a_sNn_{eq}\eta-a^4_{z}=0
\;.\end{equation}
Our results are shown in Fig. 1 and in Fig. 2. The former figure corresponds to $\gamma=5/3$, the latter to $\gamma=2$. The left part of these two figures represent the cigar-shaped case (CS), while the right part the disk-shaped (DS) one. In both cases ($\gamma=5/3$, $\gamma=2$), the left-top panel shows the atomic cloud radial width $\sigma$ as a function of the equilibrium density $n_{eq}$, while the right-top panel shows the axial width $\eta$ studied by varying the density. From such panels, one can see that when the density is sufficiently low, $\sigma$ and $\eta$ give back, respectively, the radial - $a_{\bot}$ -  and axial - $a_z$ - characteristic harmonic oscillator lengths.

Both for $\gamma=5/3$ and $\gamma=2$, in the left-bottom and right-bottom we show the behavior of the sound velocity $c_s$ as a function of the equilibrium density. The solid lines have been obtained by Eqs. (\ref{3d1dc1}), (\ref{3d2dc1}), (\ref{3d1dc2}), and (\ref{3d2dc2}), with the atomic cloud widths ($\sigma$ for the cigar, and $\eta$ for the disk) as functions of the density found by numerically solving, respectively, Eqs. (\ref{3d1dw1}),(\ref{3d2dw1}),(\ref{3d1dw2}), and (\ref{3d2dw2}). The dotted lines represent the sound velocity in the pure 3D regime. This has been obtained by  Eqs. (\ref{3d1dc1}), (\ref{3d2dc1}), (\ref{3d1dc2}), and (\ref{3d2dc2}); for this case the cloud widths in terms of density is given by the asymptotic behavior ($n_{eq}\rightarrow \infty$) of the solutions of Eqs.(\ref{3d1dw1}),(\ref{3d2dw1}),(\ref{3d1dw2}), and (\ref{3d2dw2}), respectively. Finally, the dotted lines have been obtained by solving Eqs. (\ref{3d1dc1}) and (\ref{3d1dc2}) with $\sigma=a_{\bot}$, that is the 1D case, and Eqs. (\ref{3d2dc1}) and (\ref{3d2dc2}) with $\eta=a_{z}$, that is the 2D case. Then, from the left-bottom and right-bottom panels of Figs. 1-2, it can be seen that our numerical solutions  (solid lines) approximates very well the results of the 1D (2D) regime (dashed lines) for sufficiently low densities, while the full 3D regime behavior (dotted lines) is well described by our solutions for high densities. So our solutions capture the physics both in the extreme regimes and in the intermediate region.
In determining the properties of the system, the density plays an important role. In fact, when this latter quantity is low, the particles experience few interactions and therefore the dynamics takes place mainly where there the harmonic confinement is absent. On the other hand, in the presence of sufficiently high densities the interactions between the particles are enhanced. This implies that the previously forbidden directions (i.e. those where there is the harmonic trap) begin to be involved in the dynamics.

As a conclusive remark, we observe that it is possible to study the sound velocity behavior by following another approach as well.  We can determine, in fact, analytically $n_{eq}(\sigma)$ and $n_{eq}(\eta)$ (instead of $\sigma(n_{eq})$ and $\eta(n_{eq})$) from Eq. (\ref{sigmaconst}) and Eq. (\ref{etaconst}), respectively, that employed in Eq. (\ref{soundcigar}) and Eq. (\ref{sounddisk}) give
\begin{equation}
\label{soundcigar2}
c_{s}=\omega_{\perp}\left(\frac{\gamma(\sigma^{4}-\lambda a^{4}_{\perp})}{2\sigma^{2}}\right)^{\frac{1}{2}}
\end{equation}
in the cigar-shaped case, and
\begin{equation}
\label{soundisk2}
c_{s}=\omega_{z}\left(\frac{\gamma(\eta^{4}-\lambda a^{4}_{z})}{2\eta^{2}}\right)^{\frac{1}{2}}
\end{equation}
for the disk-shaped configuration.

These formulas could have a greater utility in experiments rather than a theorical use, because they allow to determine the wave propagation velocity by simply measuring the average width of the superfluid confined gas.

\section{Conclusions}
We have considered a confined two-component Fermi gas, at zero temperature, both in the cigar- and in the disk-shaped configuration realized by a strong harmonic potential in the radial plane and in the axial direction, respectively. We have studied this system via the Popov Lagrangian density by employing a Gaussian ansatz on the density. By integrating out the directions where the dynamics is frozen, a dimensional reduced Lagrangian has been achieved. We have studied the associated Euler-Lagrange equations by supposing to create a sufficiently weak perturbation with respect to the equilibrium when the system is in the stationary regime. We have numerically determined the behavior of the sound velocity as a function of the equilibrium density, both in the BCS side and in the BEC region of the crossover, that has been compared with that of the pure three-dimensional (3D) regime and to those corresponding to the pure one-dimensional (1D) (for the cigar-shaped case) and two-dimensional (2D) (for the disk-shaped case) regimes. Thanks to this comparison, we have pointed out that our numerical solutions well approximates the 3D results for sufficiently high densities and the 1D (2D) behavior when the density is low enough, our numerical solutions being able to capture the physics in the intermediate regime as well.\\

This work has been supported by MIUR (PRIN 2010LLKJBX). GM and LS acknowledge financial support from
the University of Padova (Progetto di Ateneo 2011) and Cariparo Foundation (Progetto di Eccellenza 2011). GM acknowledges financial support also from Progetto Giovani 2011
of University of Padova.

\end{document}